\documentclass[12pt]{article}
\usepackage{epsfig}

\def\beq{\begin{equation}}
\def\eeq#1{\label{#1}\end{equation}}
\def\eeqn{\end{equation}}

\def\beqa{\begin{eqnarray}}
\def\eeqa#1{\label{#1}\end{eqnarray}}
\def\eeqan{\end{eqnarray}}

\let\bar=\overbar

\def\Dslash{\not{\hbox{\kern-4pt $D$}}}
\def\dslash{\not{\hbox{\kern-2pt $\del$}}}

\def\msb{{\bar{\ssstyle M \kern -1pt S}}}

\usepackage{fancyhdr,graphicx}
\def\Title#1{\begin{center} {\Large {\bf #1} } \end{center}}

\begin{document}

\Title{Gravitational Wave Generation in Rotating Compact Stars}

\bigskip\bigskip

%+\addcontentsline{toc}{chapter}{{\it L. Skywalker}}
%+\label{SkywalkerLukeStart}

\begin{raggedright}

{\it Ana Maria Endler and S\'ergio Barbosa Duarte \\
Centro Brasileiro de Pesquisas F\'isicas, 
Rua Dr. Xavier Sigaud 150, 
Urca, Rio de Janeiro, 
22290-180, Brazil\\
{\tt Email:endler@cbpf.br}}

\bigskip

{\it Hil\'ario Rodrigues\\
Centro Federal de Educa\c{c}\~ao Tecnol\'ogica do Rio de Janeiro, 
Av. Maracan\~a 229, 
Maracan\~a, Rio de Janeiro, 
20271-110, Brazil
}

\bigskip

{\it Marcelo Chiapparini\\
Instituto de F\'{\i}sica, Universidade do Estado do Rio de Janeiro, 
Rua S\~ao Francisco Xavier 524, 
Maracan\~a, Rio de Janeiro, 
20550-900, Brazil
}

%\bigskip\bigskip
\end{raggedright}

\section{Introduction}

 According to the General Theory of Relativity deformed rotating neutron stars irradiate gravitational waves. The energy carried away by gravitational radiation changes the rotation state, and the compact object evolves to a new equilibrium configuration.  
In this work we present a simplified description of this phase, describing the system by an  uniformly rotating triaxial homogeneous ellipsoid to catch the main aspects of the evolution. We construct an effective Lagrangian model, in which the kinetic energy associated to the breath mode and rotation are explicitly determined. The rate of gravitational waves radiation is determined in the framework of the weak field limit approximation of Einstein equations. We then solve numerically the equations of motion for the nascent neutron star, incorporating the diffusion of neutrinos in the calculation.

\section{Lagrangian model}

The dynamics of an uniformly rotating neutron star is approximated by a compressible homogeneous triaxial ellipsoid.  We construct an effective Lagrangian
\begin{equation}
{\cal L} = K  - W - U_{rot} - U_{int},
\end{equation}
with $K$ being the translational kinetic energy, $W$ the gravitational potential energy, $U_{int}$ the internal energy, and $U_{rot}$ the rotational kinetic energy.  All quantities are written in terms of the semi-axes of the ellipsoid $a_1$, $a_2$, and $a_3$, and their respective time derivatives. 

The Newtonian gravitational energy of the triaxial ellipsoid is given by \cite{Chandra}
\begin{equation}
W=-\frac{3}{10}GM^{2}\frac{A}{a_{1}a_{2}a_{3}},  \label{19}
\end{equation}%
where $A$ is defined by 
\begin{equation}
A=\sum_{i=1}^{3}A_{i}a_{i}^{2}, \, \, \, \, i=1,2,3 ,\label{4}
\end{equation}%
where 
\begin{equation}
A_{i}=a_{1}a_{2}a_{3}\int_{0}^{\infty }\frac{d\zeta }{\Delta \left(
a_{i}^{2}+\zeta \right) },  \label{3}
\end{equation}%
with
\begin{equation}
\Delta ^{2}=(a_{1}^{2}+\zeta )(a_{2}^{2}+\zeta )(a_{3}^{2}+\zeta ).
\label{5}
\end{equation}

The translational kinetic energy associated with the internal motion 
of the ellipsoid is given by the simple quadratic form
\begin{equation}
K=\frac{1}{10}M\left(
\dot{a}_{1}^{2}+\dot{a}_{2}^{2}+\dot{a}_{3}^{2}\right) .
\label{23}
\end{equation}

The rotational energy measured in the rest frame fixed at the center of the ellipsoid is given by 
\begin{equation}
U_{rot}=\frac{1}{2}I\Omega ^{2},  \label{24}
\end{equation}%
where $I$ is the moment of inertia of the ellipsoid relative to the axis of rotation, which reads%
\begin{equation}
I=\frac{1}{5}M(a_{1}^{2}+a_{2}^{2}).  \label{25}
\end{equation}

The angular velocity can be expressed in terms of the total
angular momentum $L$ of the ellipsoid, $ L=I \Omega$, and thus we may put the rotational energy
(\ref{24}) in the form
\begin{equation}
U_{rot}=\frac{5}{2}\frac{L^{2}}{M(a_{1}^{2}+a_{2}^{2})}.
\label{27}
\end{equation}

The equations of motion for the three semi-axes obtained from the Lagrangian of the system are thus given by
\begin{equation}
\frac{d^2a_1}{dt^2} = - \frac{3}{2}\frac{GM}{a_{2}a_{3}}A_{1}+\frac{25 L^{2}}{M^{2}}%
\frac{a_{1}}{(a_{1}^{2}+a_{2}^{2})^{2}}+\frac{20\pi }{3M}Pa_{2}a_{3}, \label{29}
\end{equation}
\begin{equation} 
\frac{d^2a_2}{dt^2} = - \frac{3}{2}\frac{GM}{a_{1}a_{3}}A_{2}+\frac{25L^{2}}{M^{2}}%
\frac{a_{2}}{(a_{1}^{2}+a_{2}^{2})^{2}}+\frac{20\pi }{3M}Pa_{1}a_{3},
\label{30}
\end{equation}
and
\begin{equation}
\frac{d^2a_3}{dt^2} = - \frac{3}{2}\frac{GM}{a_{1}a_{2}}A_{3}+\frac{20\pi }{3M}%
Pa_{1}a_{2},  \label{31}
\end{equation}
where $P$ is the fluid pressure. 

\section{Gravitational wave radiation and neutrino escaping}
 
Within the weak field approximation, the time rate of the system energy carried out by the gravitational waves is provided by the formula \cite{Landau,Ohanian}
\begin{equation}
 \frac{dE}{dt} = -\frac{G}{45c^{5}} \left( \frac{\partial ^{3}Q_{\alpha \beta }}{
\partial t^{3}}\right)^{2},  \label{52}
\end{equation}
from which we can obtain the mean value of the luminosity 
\begin{equation}
L_{\rm GW} =-\left\langle \frac{dE}{dt} \right\rangle.  \label{52d}
\end{equation}

For a homogeneous triaxial ellipsoid rotating uniformly about the $x_{3}$-axis with the frequency $\Omega$, the luminosity is  \cite{Beltrami}:
\begin{equation}
L_{\rm GW} = \frac{32}{125}\frac{GM^{2}\Omega ^{6}}{c^{5}}\left(
a_{1}^{2}-a_{2}^{2}\right) ^{2}.  \label{61}
\end{equation}

The rate of angular momentum loss due to the gravitational radiation is also determined as:% \cite{Chau}:
\begin{equation}
\frac{dL}{dt}=-\frac{32}{125}\frac{GM^{2}\Omega ^{5}}{c^{5}}\left(
 a_{1}^{2}-a_{2}^{2}\right) ^{2}.  \label{63}
\end{equation}

We assume for $t = 0$ that protons, neutrons, electrons and trapped neutrinos are in beta equilibrium, such that
\begin{equation}
\mu_n(0)  =  \mu_p(0) + \mu_e(0) - \mu_\nu(0) \label{beta}.
\end{equation} 

 The electric charge neutrality and baryon number conservation lead to the additional constraints: $n_e(0) = n_p(0)$ and $\rho(0) = n_n(0) + n_e(0)$, where  $n_{e} (0)$, $n_{p} (0)$, $n_{n} (0)$ and $\rho(0)$ are the initial electron density, proton density, neutron density and baryon density, respectively. Moreover, the neutrino trapping at $t = 0$ yields the relation $n_\nu (0) = Y_{L_e} (0) \rho(0) - n_e(0),$ where $Y_{L_e} (0)$ is the initial leptonic fraction (we adopt here the value $Y_{L_e}(0) = 0.4$). These constraints, including the Eq. (\ref{beta}), provide the initial neutrino density as a function of the initial baryon density.      

In order to mimic the neutrino emission, we assume that neutrinos leave the system at a constant rate, given by
\begin{equation}
\frac{d N_{\nu}}{dt} = - \tau ,
\end{equation}
where $N_{\nu}$ is the number of neutrinos trapped in the system at the time $t$, and $ \tau $ is a characteristic constant. So, integrating the last equation, we get 
\begin{equation}
n_{\nu} (t) = \frac{\rho (t)}{\rho (0)} n_{\nu} (0) e^{- t/\tau},  
\end{equation}
where  $\rho(t)$ is the baryon density at the time $t$.  With the escape of 
 neutrinos the pressure reduces, and thus the system becomes unstable against gravity.

The chemical potential of the degenerate neutrinos is $\mu_{\nu} = \left( 6 \pi^2 n_\nu \right)^{1/3}$, which contributes with the partial pressure $P_\nu = \left( 1 / 24 \pi^2 \right) \mu_\nu^4$.

\section{Results and conclusions}

We assume that the proto-neutron star is initially composed of baryonic matter, and leptons. The equation of state in the hadronic regime considers the
contribution of the complete baryonic octet in a relativistic mean
field approach \cite{glen}. This equation of state is connected to the bag
model equation of state for high barionic density, in the deconfined
quark regime. In the low density regime of interest, the BPS \cite{bps} 
equation of state is used as a smooth extrapolation.

We solve numerically the equations of motion (\ref{29})--(\ref{31}) for a given initial condition. The semi-axes of the ellipsoid, as well as the initial angular frequency, are determined for the mass $M=1.6$ $M_{\odot}$, provided the initial angular momentum $L_0$. 

We consider in this work two different values of the initial angular momentum. For $ L_{0}=1.8 \times 10^{49} \ {\rm g}\cdot {\rm cm}^{2} \cdot {\rm s}^{-1} $ we obtain the initial values for the semi-axes of the ellipsoid $a_{1}(0) = 2.05\times 10^{6}\ {\rm cm}$, $a_{2}(0) =1.59\times 10^{6}\ {\rm cm} $ and $a_{3}(0) = 1.01\times 10^{6}\ {\rm cm}$. For this initial condition, the obtained values of the initial baryon density and neutrino density are $\rho(0) = 0.14 \ {\rm fm}^{-3} $ and $n_{\nu }(0) =1.40\times 10^{-2}\ {\rm fm}^{-3}$, respectively. 

For $ L_{0}=2.6 \times 10^{49} \ {\rm g}\cdot {\rm cm}^{2} \cdot {\rm s}^{-1} $ the initial values of the semi-axes are $a_{1}(0) = 2.96 \times 10^{6}\ {\rm cm}$, $a_{2}(0) =1.48 \times 10^{6}\ {\rm cm} $ and $a_{3}(0) = 8.6 \times 10^{5}\ {\rm cm}$. In this case the values of the initial baryon density and neutrino density are $\rho(0) = 0.12 \ {\rm fm}^{-3} $ and $n_{\nu }(0) =1.24 \times 10^{-2}\ {\rm fm}^{-3}$, respectively.

%%%%%%%%%%%%%%%%%%%%%%%%%%%%%%%%%%%%%%%%%%%%%%%%%%%%%%%%%%%%%%%%%%%%%%%%%
%%
%%   use this format to include an .eps figure into your paper
%%
\begin{figure}[htb]
\begin{center}
\includegraphics[width=13cm]{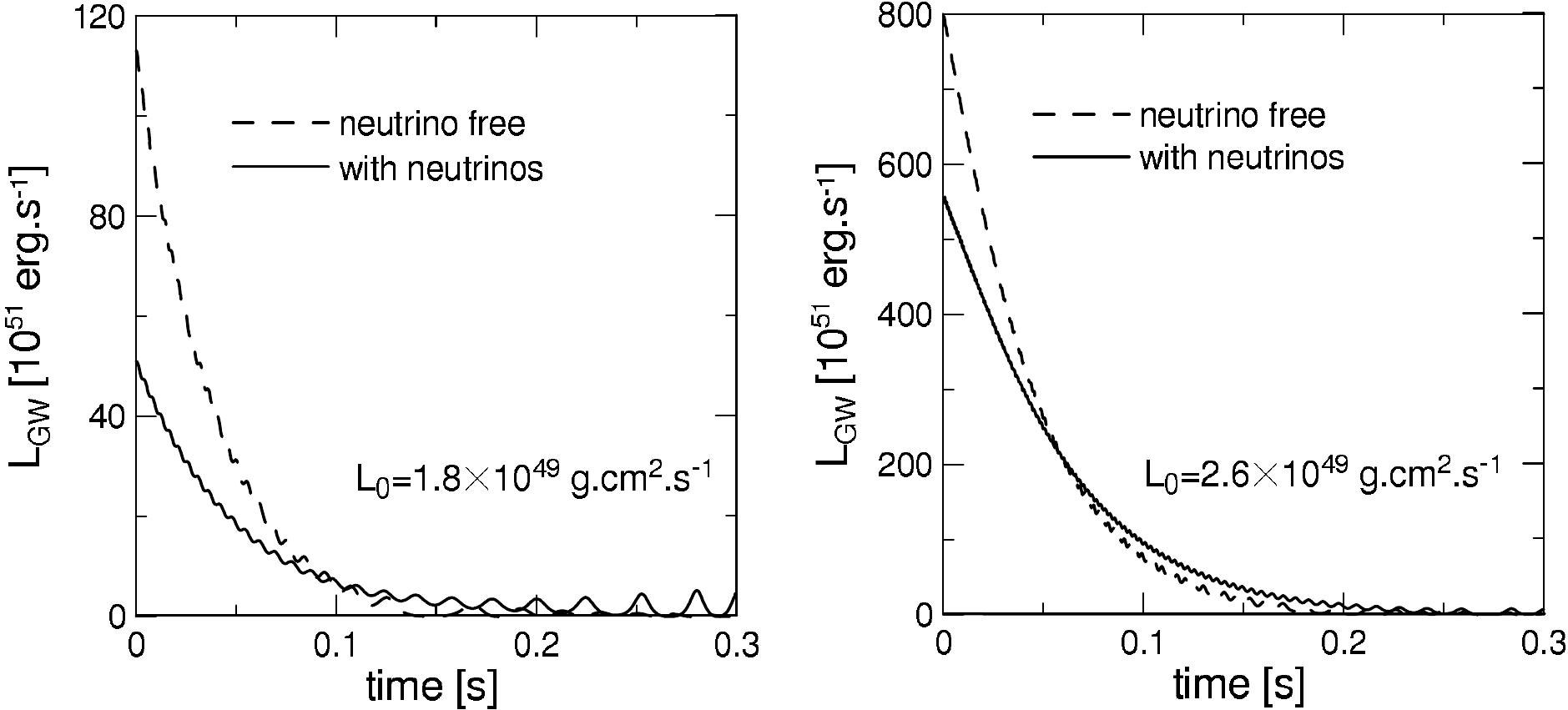}
\caption{Gravitational wave luminosity as a function of time, for two values of the initial angular momentum $L_0$. The used neutrino diffusion parameter is $\tau = 10$ s.}
\label{fig:Fig1}
\end{center}
\end{figure}
%%%%%%%%%%%%%%%%%%%%%%%%%%%%%%%%%%%%%%%%%%%%%%%%%%%%%%%%%%%%%%%%%%%%%%%%%%%

In Figure \ref{fig:Fig1} we depict the gravitational luminosity as a function of time for two different values of the initial angular  momentum $L_0$. We see that the luminosity decreases rapidly from hundreds to few tenths of $10^{51} \, {\rm erg/s}$. Notice the effect of the neutrino diffusion on the initial luminosity of the gravitational wave, compared with the case where neutrinos, are absent.  The obtained final states are uniformly rotating objects possessing period, mean radius, density and chemical composition typical of neutron stars.  Neutrino diffusion seems to play an important role in the gravitational wave radiated by proton-neutron stars.

%\section*{Acknowledgments}

%H. Rodrigues and S. B. Duarte thank CNPq
%for financial support.

\end{document}